\documentstyle[aps,multicol,prl]{revtex}
\renewcommand{\narrowtext}{\begin{multicols}{2} \global\columnwidth20.5pc}
\renewcommand{\widetext}{\end{multicols} \global\columnwidth42.5pc}
\begin{document}

\draft
\narrowtext
\begin{flushleft}
{\large\bf Comment on: ''Different Results for the Equilibrium Phases
of Cerium above 5\ GPa``}
\end{flushleft}

\bigskip
A recent Letter by McMahon and Nelmes\ \cite{McMNel} claims that the
dispute on the equilibrium phases of cerium metal beyond 5\ GPa has
arisen simply from differences in the sample production and preparation.
Our previously published results on the $4f$\/ occupation numbers of
cerium metal under high pressure\ \cite{RoeKri} contradict the essential
of their claim that cerium between 5 and 12\ GPa has a stable single
phased equilibrium structure (either monoclinic C2/{\it m} {\it or}
orthorhombic $\alpha$-U). Actually, we have put forward strong evidence
for the coexistence of at least two structurally different phases to be
near the true equilibrium situation across the entire $\alpha$',
$\alpha$'' phases\ \cite{RoeKri}.

First, the thermodynamic equilibrium of the Ce phases is strongly
influenced by the electronic instability of cerium in the solid state
depending on temperature and volume, in particular on the $4f$\/
occupation number, $\nu$, \cite{RoeKri}, and on the volume dependence of
the $4f$\/ hybridization \cite{AllMar}. We have performed measurements
of $\nu$ (closely related with the fractional valence, $v=3+\nu$) in
elemental cerium under pressures up to 12\ GPa. The valence is near 3 at
ambient pressure, jumps to 3.14 in the $\gamma-\alpha$ transition at
0.8\ GPa, increases to 3.26 near 3\ GPa. Most important, beyond 5 GPa it
remains constant at 3.22 across the entire $\alpha$', $\alpha$'' phases\
\cite{RoeKri}. The saturation of the valence in elemental Ce between 5
and 12\ GPa corroborates the experimental results obtained earlier from
numerous x-ray absorption and photoemission studies of so-called
$\alpha$-like intermetallic compounds and dilute alloys yielding Ce
valences clustering near 3.25 \cite{RoeKri}. We have shown \cite{RoeKri}
that the saturation of the valence beyond 5 GPa finds a straightforward
thermodynamic explanation in the formation of two or more coexisting
structural phases. From a Maxwell construction of the two phase region
we have been able to obtain the additional degree of freedom accounting
for the pressure independent (average) fractional valence in the range
5--12 GPa.

Second, about 20 years ago Zachariasen \cite{Zac} pointed to the effects
of sample production and preparation on the structures of Ce between 5
and 12\ GPa. From his seminal studies of the high-pressure forms of
elemental cerium he concluded ''it is not known at the present time why
the transformation at about 51 kbar sometimes leads to the formation of
the stable phase $\alpha$'-Ce and at other times to the formation of the
metastable phases $\alpha$''-Ce I or $\alpha$''-Ce II, or to the formation
of a mixture of $\alpha$'-Ce and $\alpha$''-Ce phases. Small impurities,
rate of change of pressure, thermal and pressure history of the sample,
and anisotropy of the applied pressure are possible factors that could
influence the nature of the transition``. McMahon and Nelmes
\cite{McMNel} believe that the answer to this problem is ''simply`` in
the difference between ''cold-worked non-USA cerium`` (filings) yielding
only one monoclinic form (not two: $\alpha$''-Ce I and $\alpha$''-Ce II),
and ''non-cold-worked USA cerium`` (slices) yielding only the
orthorhombic form $\alpha$'-Ce ($\alpha$\/-U structure). Although the
chemical, thermal, and mechanical history of the samples has not been
not under the full control of McMahon and Nelmes \cite{McMNel}, they
state the origin of the structural differences to be in the locus of the
sample production and in ''cold-working``. It is well established that
slices cut from an ingot of Ce are exerted to ill defined strain
transforming them at least partially into the $\alpha$\/-phase. Thus
produced slices have to be considered as beeing ''cold-worked`` as well
as the filings. However, the effect of surface tension is different for
the (unannealed) small grains (filings) transforming much easier into an
$\alpha$\/-like state than for the larger slices tending to stabilize a
$\gamma$\/-like state. The effect of surface tension on the structural
and electronic properties (e.g. lattice parameters, magnetic
susceptibility, $4f$\/ occupation number) shows up already at ambient
pressure and may be used to define 
properly the initial conditions of the high pressure experiment.

In conclusion, the statement of McMahon and Nelmes\ \cite{McMNel} that
the C2/$m$ structure is the only monoclinic form of Ce in the range
5--12\ GPa appears to be unfounded from their experimental procedures.
To establish the structural equilibrium situation beyond 5\ GPa a
detailed and complete control of all parameters that could influence the
equilibrium is required. Sample characterizations of the type
 ``cold-worked non-USA'' and ``non-cold-worked USA'' are misleading.
\begin{flushleft}
\large{J. R\"ohler}
\begin{quote}
\small{
Universit\"at zu K\"oln
\\II. Physikalisches Institut
\\Z\"ulpicherstr. 77, D-50937 K\"oln
\\Germany
\\abb12@rs1.rrz.uni-koeln.de
}
\end{quote}
\smallskip
\small{
Received
\\PACS numbers: 61.55.F, 62.50, 74.70.S
}
\end{flushleft}

\widetext

\begin{references}
\bibitem{McMNel}M.I. McMahon, and R.J. Nelmes, Phys. Rev. Lett. {\bf 78},
3884 (1997).

\bibitem{RoeKri}J. R\"ohler, D. Wohlleben, J.P. Kappler, and G. Krill,
Phys. Lett. {\bf 103A}, 220 (1984); D. Wohlleben, and J. R\"ohler, J.
Appl. Phys. {\bf 55}, 1904 (1984).

\bibitem{AllMar}J.W. Allen, R.M. Martin, Phys. Rev. Lett. {\bf 49}, 1106
(1982).

\bibitem{Zac}W. H. Zachariasen, Proc. Natl. Acad. Sci. USA {\bf 75},
1066 (1978).

\end{references}
\end{document}